\documentclass[aps,prd]{revtex4}
\usepackage{graphicx}
\newcommand{\beq}{\begin{eqnarray}}
 \newcommand{\eeq}{\end{eqnarray}}
\newcommand{\be}{\begin{equation}}
 \newcommand{\ee}{\end{equation}}

\def\fun#1#2{\lower3.6pt\vbox{\baselineskip0pt\lineskip.9pt
\ialign{$\mathsurround=0pt#1\hfil ##\hfil$\crcr#2\crcr\sim\crcr}}}

\newcommand{{\SD}}{\rm SD}
\newcommand{{\GeV}}{\rm GeV}
\newcommand{{\fm}}{\rm fm}
\newcommand{{\oMS}}{\overline{MS}}

\newcommand{\vex}{\mbox{\boldmath${\rm x}$}}

\newcommand{\ver}{\mbox{\boldmath${\rm r}$}}
\newcommand{\vesig}{\mbox{\boldmath${\rm \sigma}$}}

\newcommand{\vep}{\mbox{\boldmath${\rm p}$}}

\newcommand{\ven}{\mbox{\boldmath${\rm n}$}}

\newcommand{\lan}{\langle}
\newcommand{\ran}{\rangle}

\begin{document}

\title{The coupled-channel analysis  of the $D$ and $D_s$ mesons}
\date{\today}
 \author{Yu. A.~Simonov}
\affiliation{State Research Center, Institute of Theoretical and
Experimental Physics, Moscow, 117218 Russia}
\author{J. A.~Tjon}
\affiliation{Jefferson Lab., 12000 Jefferson Ave., Newport News, VA
23606, USA}
\affiliation{KVI, University of Groningen, Groningen,
The Netherlands}

\begin{abstract}

The shift of the $p$-wave  $D_s$ meson mass due to coupling to the
DK channel is calculated without fitting parameters using the chiral
Lagrangian. As a result the original $Q\bar q$ mass 2.490 MeV
generically calculated in the relativistic quark models is shifted
down to the experimental value 2317 MeV. With the same Lagrangian
the shift of the radial excited $1^-$ level is much smaller, while
the total width $\Gamma > 100$ MeV and the width ratio is in
contradiction with  the $D^*(2632)$ state observed by SELEX group.
\end{abstract}
\maketitle

\section{Introduction}

The heavy-light $D$ and $D_s$ mesons have extensively been
investigated experimentally in the last 20 years \cite{1}. Recently
a lot of attention has  been paid to the $0^+$ state of the $D_s$
meson. It was found by the BABAR group \cite{2} at the mass
$M=2318\pm 1.3 $ MeV and confirmed by CLEO and Belle \cite{3}. This
value of mass is $\sim 40$ MeV below the KD threshold and the width
is very small. On the theoretical side this state was the object of
intensive study (see Ref.~\cite{4} for reviews and references).
There exists some disparity between the theoretical predictions of
the $D^*_s(0^+)$ mass in the Relativistic Quark Model (RQM)
calculations \cite{5}-\cite{10} and experimental results. Indeed the
RQM predictions vary from 2380 MeV \cite{6,7} to 2487 MeV \cite{10},
being however substantially larger than measured experimentally.

Most recently a new $D_s$ state was observed in the SELEX experiment
\cite{14} at the mass value $M=2632$ MeV. It is argued in
Ref.~\cite{15} that this state may be associated with the radial
excited $1^{-*}$ level of $D_s$ meson, which is shifted down due to
coupling to the $\eta D_s$ and $KD$ channels. As one will see both
levels $D^*_s(2317) $ and $D_s^*(2632) $ are connected to thresholds
via chiral decays and we shall treat them  below using the chiral
Lagrangian containing only $f_{\pi},~f_K,~f_{\eta}$ as parameters.

It is the purpose of the  present paper to study the effect of
nearby thresholds on the position of resonances first in the most
general setting and then to calculate  numerically using the chiral
quark Lagrangian  without fitting  parameters. The paper is
organized as follows. In section 2 we present a general discussion
of channel coupling and level shift using the (relativistic)
Hamiltonian formalism, where we also give a classification of
possible $S$-matrix poles. In section 3 the chiral quark Lagrangian
is written down  and used to describe the decay transition $D^*_s\to
DK,~D_s \eta$. The final equation for the resonance position with
account of this decay is explicitly written. In section 4 the
numerical solution of this equation is described and the final
results are presented. The paper closes with a discussion and
comparison with other results.

\section{Resonance states in the coupled-channel system}

The relativistic quark model has been remarkably successful in
predicting the $D$ and $D_s$ meson spectrum, apart from some
exceptions of a few resonances, which are experimentally found at
substantially lower masses. Similar results have been found recently
in the relativistic Hamiltonian approach\cite{11} derived on the
basis of Field Correlator Method (FCM) \cite{12} and applied to the
$D,D_s$ mesons in Ref.~\cite{13}. The results for the masses in the
FCM analysis are shown in Table 1 and Table 2. (The entries given in
the tables are recalculated for parameters given in the Table
captions).

\vspace{1.0cm}
\newpage

 \begin{tabular}{lp{11cm}}
{\bf Table 1}&  Masses of $L=0,1$ states of $D$ mesons. Input
parameters used in the FCM calculations: $\alpha_s=0.46,
\sigma=0.17$ GeV$^2,$~ $m_c=1.44.$ GeV, $ m_n =7$
MeV\\
\end{tabular}\\

\begin{tabular}{|l|l|l|l|l|l|l|l|} \hline
 State $J^P$ & $0^-$ & $1^-$ & $0^+$ & $1^+(l)$ & $1^+(h)$ &
 $2^+$&$1^{-*}$
 \\
 & &  &  &  &  &&  \\
 \hline
Mass (MeV)  & &  &  &  &  & & \\ from \cite{13} & 1859  &2047 &
2370 & 2425 & 2455 & 2456 &2729 \\
\hline
  Mass (MeV)&  &  &  &  &  &&  \\
experiment & 1869 & 2010 & 2300$\pm60$ & 2400 & 2422 & 2459 &
(2640?)\\
\hline
$\Gamma$ (MeV) & & & & & &&\\
experiment   & ~~--- & $<0.13$ & 280 & $\sim 250$ & 20 & 23$\div$45&
$<15$\\
\hline
\end{tabular}

\vspace{1cm}

As is seen from Tables 1 and 2 the overall agreement is reasonably
good except for a few states. In particular, the $D^*_s(0^+)$ state
is one example of such a discrepancy in the prediction of
theoretical models and which can be associated with the KD threshold
at 2366 MeV.

\vspace{1cm}

 \begin{tabular}{lp{11cm}}
{\bf Table 2}&  Masses of $L=0,1$ states of $D_s$ mesons. Input
parameter used in the FCM calculations: $\alpha_s=0.46, \sigma=0.17$
GeV$^2,$~ $m_c=1.44.$ GeV, $ m_s
=0.175$ GeV\\
\end{tabular}\\

\begin{tabular}{|l|l|l|l|l|l|l|l|} \hline
 State $J^P$  & $0^-$ & $1^-$ & $0^+$ & $1^+(l)$ & $1^+(h)$ &
 $2^+$&$1^{-*}$
 \\& &  &  &  &&  &  \\
 \hline
Mass (MeV)  & &  &  &  &  &&  \\ from \cite{13} & 1929  &2087 &
2404 & 2462 & 2488 & 2494 &2774 \\
\hline
  Mass (MeV)&  &&  &  &  &  &  \\
experiment & 1968 & 2112 & 2317 & 2462 & 2536 & 2572& 2632(?) \\
\hline
$\Gamma$ (MeV) &~~--- &$<1.9$  & $<10$ & $<6.6$ &$<2.3$ & $\sim
15$&$<17$
\\ experiment& & & &$8.6\div 0.4$  &&  &  \\ \hline
\end{tabular}\\

\vspace{1cm}

For comparison in Table 3 a summary of the results is given of other
theoretical quark model predictions for this state.  A look at the
table tells us that all the theoretical predictions are about $\sim
90\div 190$ MeV higher. So one needs a shift of about this value to
get agreement with the experimental value. A similar descrepancy can
be seen from Table 2 for the $D_s^*(2632)$ resonance. We explore in
this paper whether this disparity can be explained due to the
presence of coupled channels with nearby thresholds.

\vspace{1cm}

 \begin{tabular}{lp{11cm}}
{\bf Table 3}&  Theoretical predictions of the mass $D^*_s(0^+)$
in various quark models\\
\end{tabular}\\

\begin{tabular}{|l|l|l|l|l|l|l|} \hline
Ref. & [5] &[6]&[7]&[8]&[9]&[10]\\ & &  &  &  &  &  \\
\hline
Mass (MeV)  & 2480 &2388&2380  &2508  &2455  &2487
\\
\hline
\end{tabular}\\

\vspace{1cm}

Resonances in the coupled-channel system were considered in numerous
papers both in nonrelativistic nuclear physics and in the
relativistic Hamiltonian dynamics, see \cite{16} for a review and
references. Assuming that a local or nonlocal relativistic
Hamiltonian can be written for each channel $H_i,~i=1,2,...$ and
for the Channel Coupling (CC), $V_{ij}, ~i,j=1,2,...$ the
time-independent system of equations can be written as

\be
[(H_l-E)\delta_{ll'}+V_{ll'}] G_{l'l"}=1.\label{6}\ee For two
channels it is \be\begin{array}{cc}
 ( H_1-E) G_{11} +V_{12} G_{21} =1, & ( H_1-E) G_{12} +V_{12} G_{22} =0, \\
 ( H_2-E) G_{22} +V_{21} G_{12} =1, & V_{21} G_{11}+( H_2-E)
 G_{21}=0.
\end{array}
\label{7}
\ee
The system (\ref{7}) can be reduced to the effective
one-channel problem, corresponding to the Feshbach equation
\cite{17}

\be
( H_1-E) G_{11}-V_{12}\frac{1}{H_2-E} V_{21} G_{11} =1.
\label{8}
\ee

 At this point one can classify all possible poles $E$ of
 the Green functions $G_{ik}$. These poles may originate from
 the bound states or resonances in a given channel i, located at
 $E^{(n)}_i$ and shifted due to CC to a new position, which we will denote
 by   $E^{(n)*}_i$. Another possibility  is that resonance
 poles appear solely due to the strong CC interaction -- the so-called CC
poles \cite{16,18}. These extra poles usually originate from distant
dynamical poles in the complex plane, which move close to threshold
when the CC coupling increases. The quantitative characteristics of
the CC interaction is given by the last term on the l.h.s. of
Eq.~(\ref{8}), which can be called the Feshbach potential,
\be
V_{121}(E)\equiv - V_{12} G_2 V_{21} = - V_{12} \frac{1}{H_2-E}
V_{21}.
\label{4a}
\ee
Note that $V_{121}(E)$ can support bound states
or resonances even in the case when diagonal interaction $V_i, ~~
i=1,2$ vanishes but $V_{12}=V^+_{21} $ is large enough.

Of special importance for us is the case when in one channel, e.g.
i=1, the spectrum is discrete (see Ref.~\cite{18} for a more
extensive discussion), and one is interested in the shift of the
discrete level due to the coupling to channel 2, where states can be
unconfined.

A somewhat similar approach was undertaken in recent papers
\cite{19}, where in our notations the scattering channel 2 and the
corresponding Feshbach potential $V_{212} $ was modelled to
calculate the scattering cross section in channel 2.  We shall
compare the results of Ref.~\cite{19} with ours in the concluding
section.
 Eq.~(\ref{8}) connects in general all states in
channel 1 and channel 2. If one separates one state  and neglects
all other states in channel 1, then one gets the following equation
for the position of the pole(s) in the Green function
\be
E=E_1^{(n)} - \lan n |V_{12} \frac{1}{H_2-E} V_{21} |n\ran,
\label{9}
\ee
where $E^{(n)}_1$ is the selected unperturbed level in
the channel 1. Insertion of the complete set of states $|m\ran \lan
m | $ with eigenvalues $E^{(m)}_2$ in the channel 2, yields
\be
E=E_1^{(n)} -\sum_m \lan n |V_{12}|m\ran \frac{1}{E_2^{(m)}-E} \lan
m |V_{21} |n\ran.
\label{10}
\ee
In what follows we shall be using
Eq.~(\ref{10}) to calculate the shift $\Delta E_n=E_1^{(n)*} -
E^{(n)}_1$ of the $c\bar s$ levels due to the open channel 2: $KD$
or $\eta D_s$ scattering states, neglecting interaction in these
states. The most important point is how to find the operators
$V_{12}$. In the next section we shall use the chiral Lagrangian
which will provide $V_{12}$  explicitly without free parameters.

\section{Coupled Channels  and Chiral Decays}

One  starts with the Lagrangian for the flavor SU(3) triplet of
quarks in the field of the heavy ($c$ or $b$) quark [20,21]. In
the Euclidean notations
\be
L=i\int d^4 x \psi^+ (\hat \partial +m + \hat M)\psi,
\label{1}
\ee
where the mass operator is
\be
\hat M= m\hat U = M \exp \left(i\gamma_5 \frac{2\varphi_a
t_a}{f_\pi} \right)\label{2} \ee and $t_a=\frac12 \lambda_a,
\lambda_a$ is the Gell-Mann matrix, $a=1,...8, f_\pi =0.093$ GeV
and the matrix Nambu-Goldstone SU(3) wave function is \be
\varphi_a\lambda_a= \sqrt{2}\left(
\begin{array}{ccc}
  \frac{\eta^0}{\sqrt{6}}+\frac{\pi^0}{\sqrt{2}}, & \pi^+, & K^+ \\
  \pi^- &\frac{\eta^0}{\sqrt{6}}-\frac{\pi^0}{\sqrt{2}}, & K^0 \\
  K^-, & \bar K_0, &-\frac{2\eta}{\sqrt{6}}
\end{array}\right).
\label{3}
\ee
$M$ is the (nonlocal) effective mass operator, which
 in the local  limit has the form  (see [22,23] for discussion and derivation)
\be
M=\sigma|\ver |,
\label{4}
\ee
where $|\ver|$ is the distance from
the light quark ($u,d,s,$) to the heavy quark ($c$ or $b$). Thus
the Lagrangian (7) contains effects of both confinement and chiral
symmetry breaking.

From Eqs.~(\ref{1}) and (\ref{2}), expanding the exponent in
Eq.~(\ref{2}), one can  derive the meson emission part of
Lagrangian, \be \Delta L= -\int \psi^+ (x) \sigma |\vex | \gamma_5
\frac{\varphi_a\lambda_a}{f_\pi} \psi d^4 x. \label{5} \ee This
Lagrangian can be expressed as in Ref.~[20] in terms of the standard
Weinberg Lagrangian [24]. It was used in Ref.~[25] to calculate the
decay widths of heavy-light mesons with good accuracy.

It is clear that the Lagrangian (\ref{1}) generates (due to the
various Fock components in $\psi^+$ or $\psi$) in general a
many-channel system of equations for the Green functions. It
contains  the main channel (e.g. the $D_s$ channel) and in
addition the channel(s) for its virtual decay products like the
($D+K$) channel or $(D_s+\eta)$ channel.

 In what follows we shall be working with  Eq. (\ref{10}) to
apply it  first  of all to the $D^*_s$(2317) state. In this case
$E^{(n)}_1$ refers to the $0^+$ level of the $D^*_s$ system, and
$E_2^{(m)}$ refers to the (continuous) energy of the system $D+K$
in the orbital $S$ state. One can neglect the $DK$ interaction in
the first approximation and write for the wave functions Dirac
equations \be |n\ran =\Psi(D^*_s) =\frac{1}{r} \left(
\begin{array}{c}
  G_n^{(1)}\Omega_{jlM}^{(1)}  \\
   i F_n^{(1)}\Omega_{jl'M}^{(1)}
\end{array}\right),
\label{11}
\ee
\be
\Omega^{(1)}_{jlM} = \Omega_{\frac12
1M_1},~~\Omega^{(1)}_{jl'M}= \Omega_{\frac12 0 M_1}
\label{12}
\ee
\be
|m\ran =\Psi(D) \frac{e^{i\vep\ver}}{\sqrt{2\varepsilon_pV_3}},
\label{13} \ee \be \Psi(D) =\frac{1}{r}\left( \begin{array}{c}
  G^{(2)}\Omega_{jlM}^{(2)}  \\
   i F^{(2)}\Omega_{jl'M}^{(2)}
\end{array}\right),
\label{14}
\ee
\be
\Omega^{(2)}_{jlM} = \Omega_{\frac12
0M_2},~~\Omega^{(2)}_{jl'M}= \Omega_{\frac12 1 M_2}.
\label{15}
\ee
Therefore the matrix elements in Eq.~(\ref{10}) are
\be
\lan
n|V_{12}|m\ran =- \int \Psi^+ (D^*_s)\sigma|\ver| \gamma_5
\frac{\sqrt{2}}{f_\pi}\Psi (D)
\frac{e^{i\vep\ver}}{\sqrt{2\varepsilon_pV_3}} d^3r\label{16}\ee \be
=\frac{\sqrt{2}\sigma}{if_\pi}\int \frac{d^3\ver}{r}
(G_n^{(1)+}F^{(2)} \Omega^+_{\frac12 1M_1} \Omega_{\frac12 1M_2} -
F_n^{(1)+ } G^{(2)} \Omega^+_{\frac12 0M_1}\Omega_{\frac12
0M_2})\frac{e^{i\vep\ver}}{\sqrt{2\varepsilon_pV_3}}
\label{17}
\ee
$$=\frac{\sqrt{2}\sigma \delta_{M_1M_2}}{if_\pi} \int \frac{\sin
pr}{p\sqrt{2\varepsilon_pV_3}} (G_n^{(1)+} F^{(2)} -
F_n^{(1)+}G^{(2)})dr.$$

Now $G^{(1)}, F^{(1)}$ and $G^{(2)}, F^{(2)}$ are solutions of the
Dirac equation
\be
\begin{array}{c}
  \frac{d F_n^{(i)}}{dr} -\frac{\kappa_i}{r} F_n^{(i)}
+ (\varepsilon^{(i)}_n- V_c(r) -m_i) G^{(i)}_n -MG_n^{(i)}=0, \\
  \frac{d G_n^{(i)}}{dr} +\frac{\kappa_i}{r} G_n^{(i)}
- (\varepsilon^{(i)}_n- V_c(r) +m_i) F^{(i)}_n  -MF_n^{(i)}=0.
\end{array}
\label{18}
\ee
Here $\kappa_1=1,~\kappa_2=-1,~M=\sigma r,~V_c(r)
=-\frac{4\alpha_s}{3r},$ and
$$ m_1= m_s= 0.15 \div 0.25 {\rm ~GeV},~m_2=0.$$
The connection between $\varepsilon^{(i)}_n$ and $E_i$ is \be
E_1^{(n)} = \varepsilon^{(1)}_n + m_c;~~ E_2^{(m)} =
\varepsilon^{(2)}_m + m_K+m_c + \frac{\vep^2}{2\tilde m_K}, \tilde
m_K = \frac{m_Km_D}{m_K+m_D}. \label{19} \ee If one neglects higher
states of the $D$ meson, the sum in Eq.(\ref{10}) can be rewritten
as
\be
E=E_1^{(n)} - \sum \frac{V_3d^3\vep}{(2\pi)^3}~ \frac{|\lan
n|V_{12}|m\ran|^2}{E^{(m)}_2(p)-E}.
\label{20}
\ee
From
Eqs.~(\ref{16}) and (\ref{20}) it is clear  that the free Green
 function  of the KD system has the form (we take into account the
 fact   that only $S$-waves  of  KD are involved)
\begin{equation}
\begin{array}{c}
G_0 (k,x,x') =\left( \int \frac{d^3p}{(2\pi)^3}
\frac{exp[i\bf{p}.(\bf{x}-\bf{x'})]}{2\omega(\vep)\left(
\frac{\bf{p}^2}{2\tilde m_K}-\Delta E\right)}\right)_{S-waves} \\
\\
=\frac{1}{4 \pi} \frac{\tilde m_K}{ \omega(k)} \frac{\sin (kx_<)
 \exp(ikx_>)}{kxx'},
\end{array}
\label{21}
  \end{equation}
where
\be
k^2=2\tilde m_K (E-m_D-m_K),~~ \omega(k) =\sqrt{k^2+ {
m}^2_K}.
\label{22}
\ee
Finally Eq.~(\ref{20}) can be rewritten as
\be
E=E^{(n)}_1 - \int V(x) V^+(x') d^3 xd^3 x'G_0 (k,x,x),
\label{23}
\ee
where  $V(x)$ is
\begin{equation}
\begin{array}{c}
  V(x)= \bar \psi_2 (x) \frac{\sqrt{2}\sigma |\vex|}{f_\pi} \gamma_5 \psi_1 (x)= \\
  =\frac{\sqrt{2}\sigma}{if_\pi x} (G^{(1)+} F^{(2)} - F^{(1)+} G^{(2)})
  \Omega^+_{\frac12 0M_1}\Omega_{\frac12 0M_2}
\end{array}
\label{23a}
\end{equation}
and we have used relation $\Omega_{\frac12 1M} = \vesig \ven
\Omega_{\frac12 0M}$. As a first approximation one can use the
fact that functions $G^{(i)}, F^{(i)}$ are concentrated around the
middle point  $x\cong b$ and write
\be
V(x) \cong  C\delta (x-b);  ~~ C= \int V(x)dx  \label{24} \ee $$
b=\frac{\int x V(x) d x}{\int  V(x) dx}.$$ As a result one obtains
\be
E= E^{(n)}_1 -C^2  \frac{4 \pi b^2 \tilde m_K  \sin (kb)
e^{ikb}}{\omega (k) k},
\label{25}
\ee
where $k=\sqrt{2 {\tilde m_K} (E-M_{D}-m_K)}$ is the relative momentum
of the K meson.

In the vicinity of the DK threshold one can replace $\omega (k)
\cong m_k$ (this is implied by  the form of Eq. (\ref{21})). Eq.
(\ref{25}) is a transcendental equation for the position of the pole
$E$. Since $C$ does not depend on k Eq. (\ref{25}) has a simple
square-root threshold at $E=m_D+ m_{K}$. The starting position of
$E=E_{pole}$ is at $E^{(n)}_1$. When one takes into account the
second term in Eq.~(\ref{25}) with gradually increasing $C^2$ the
pole moves to the final value in the upper physical k-sheet. One
expects that the trajectory will go down in mass, possibly nearby
the final value of $m(D^*_s)=2317$ MeV. In Fig.~1 we display the
trajectory of  the pole solution of Eq. (27) in $k$-plane
parameterized by $C$.
 One can see that for strong enough channel coupling the resonance
 pole moves down under the DK threshold, which will be
 substantiated by the  exact calculation of $V_{12}$, given by Eq.~(18).

\begin{figure}
\begin{center}
\includegraphics[width=10cm]{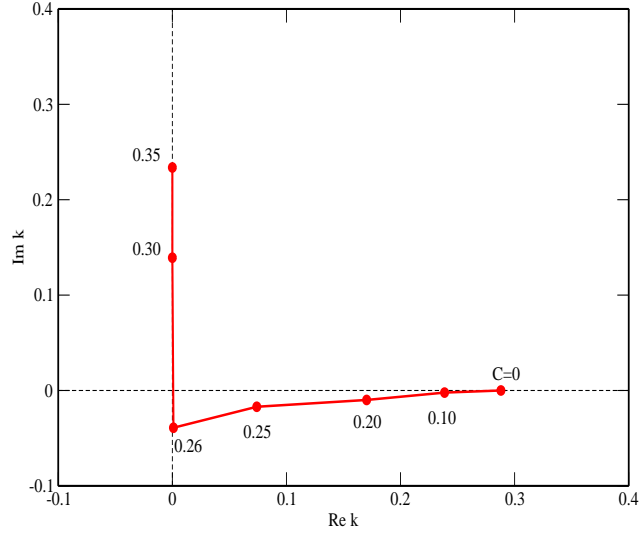}
\caption{ The trajectory of the $D_s^{*}$ pole in the complex
$k$-plane  parameterized by C. For b we have used
$b=0.4$~GeV$^{-1}$.} \hfill \label{fig1}
\end{center}
\end{figure}

We now turn to the case of the $D^*_s(2632)$ \cite{14} assuming
after Ref.~\cite{15} that it can be associated with the radially
excited $D_s^*(1^-)$ state. From Table 2 one can see that the
expected shift should be around 100--150 MeV downwards, and from the
channels $DK, D_s\eta$ the decay is in the $p$-wave. As before we
shall use Eqs. (12) and (17) where now instead of $D_s^*(0^+)$ one
should write the $D^*_s(1^-)$ state, i.e. \be \Omega^{(1)}_{jlM}
=\Omega_{\frac12 0M_1},~~ \Omega_{jl'M_1}^{(1)}=\Omega_{\frac12
1M_1}. \label{28} \ee The wave function in channel 2 is either the
same as in Eqs. (\ref{13}) -(\ref{15}), or in the case of the $\eta$
channel, one should replace $D$ by $D_s$, and the $K$ meson with
momentum $\vep$  by the $\eta$ meson with momentum $\vep'$. With
these assignments of the various states Eq.~(\ref{16}) retains its
form, but Eq.~(\ref{17}) becomes

\be
\lan n|V_{12}|m\ran =\frac{\sqrt{2} \sigma}{if_\pi}\int
\frac{d^3\ver}{r} (G_n^{(1)+}F^{(2)} \Omega^+_{\frac12 0M_1}
\Omega_{\frac12 1M_2} - F_n^{(1)+ } G^{(2)} \Omega^+_{\frac12
1M_1}\Omega_{\frac12 0M_2})
\frac{e^{i\vep\ver}}{\sqrt{2\varepsilon_pV_3}}
\label{29}
\ee
$$=-\frac{ \sqrt{2} \sigma }{f_{\pi} p\sqrt{2\varepsilon_{p}
V_3}}\int^\infty_0 dr \left( \frac{\sin p r}{p r} -\cos p r\right)
(G_n^{(1)+} F^{(2)} - F_n^{(1)+}G^{(2)}),$$
where p is the momentum of the K or $\eta$ meson.

Here the radial quantum number in Eq. (\ref{29}) is the first radial
excited state $n=1$. To compute the transition potential matrix
elements for the $D_s^*\leftrightarrow DK$ coupling one can use the
coordinate representation as in Eqs. (\ref{17})-(\ref{23a})
including Eq.~(\ref{28}), or directly calculate the $d^3\vep$
integral in Eq.~(\ref{20}). In the latter way one needs to compute
Eq.~(\ref{20}), $\omega(p)\equiv \varepsilon_p=\sqrt{p^2+m^2_K}$

\be
E=E_1^{(n)} - \frac{1}{(2\pi)^3} \int
\frac{d^3\vep|v_{12}^{(K)}(p)|^2}{2\omega
(p) (E_2(\vep)-E)}
\label{30}
\ee
 and $v_{12}(\vep)$ is
\be
v_{12}^{(K)}(\vep)=V_{12}\cdot \sqrt{2\varepsilon_p}.
\label{31}
\ee

Eq.~(30) has to be extended to also include the $D_s \eta$-channel
contribution. Observation of Eqs. (\ref{3}) and (\ref{5})  shows
that in Eq.~(\ref{3}) only the lowest diagonal term enters in the
transition potential matrix element. Hence we get for $\lan
n|V_{12}|m\ran$ an additional factor
$\left(-\frac{2}{\sqrt{6}}\right)$. As a result we have
$$v_{12}^{(\eta)}(p)=-\frac{2}{\sqrt{6}} ~v_{12}^{(K)}(p).$$
The modification of $V_{12}$, Eq.~(29), for the case of the
$\eta$-channel is straightforward. Clearly, the assignment of the
$D_s^*$ state in Eq.~(\ref{28}) remains the same and we have to
replace in Eq.~(\ref{19}) $m_K\to m_\eta$ and $m_D\to m_{D_s}$.
Moreover, the state $\Psi(D)$ in Eq.~(17) has to be replaced by
$\Psi(D_s)$, where $F^{(2)}, G^{(2)}$ now refer to the $D_s$ state.

 \section{Mass shift in  the chiral Lagrangian formalism}

We may determine the energy shift $\Delta E$ using the Dirac wave
function as found by solving Eq. (\ref{18}) for the case of an
effective quark mass operator  in the field correlator method
\cite{23}. In the leading order it has a nonlocal form and can be
parameterized as \be M(x,y)\approx~ \frac{1}{2T_g\sqrt{\pi}}
\sigma\left|\frac{\mbox{\boldmath ${\rm x}$}+\mbox{\boldmath ${\rm
y}$}}{2}\right| \exp\left(-\frac{({\bf x}-{\bf
y})^2}{4T_g^2}\right), \label{32} \ee where $\sigma$ is the string
tension and $T_g$ the gluon correlation length, characterizing the
scale of nonlocality. Note that for $T_g\to 0$ one obtains from
Eq.~(28) the local limit (10). The physical value of $T_g$ found on
the lattice  and analytically  is small, $T_g=0.25$ fm  [12].

For the case of a single coupled channel $D_s^*\leftrightarrow DK$
we may write Eq. (30) as

\be
\Delta E= - \frac{1}{(2 \pi)^3} \int
\frac{d^3\vep|v_{12}^{(K)}(p)|^2} {2 \omega(p) (E(\vep)-E_0-\Delta
E)},
\label{33}
\ee
where $\Delta E=E-E_0$ with
$E_0=m_{D_s^{*}}-m_D-m_K$  is the mass shift and $E(\vep)$  the
kinetic operator $E(\vep)=\frac{p^2}{2 {\tilde m}_K}$.

In the considered QCD string model we have taken $T_g=0.25~fm$, in
accordance with lattice gauge simulations \cite{12}. For a given
string tension $\sigma$ the wave functions of the $D$ and $D_s$
system can be found as solutions of the Schwinger-Dyson-Dirac
equation\cite{23} for the light-heavy quark system. It is given by
Eq. (19) with the quark mass operator (\ref{32}). The latter is
found from the selfconsistent solution of nonlinear equations, (Eqs.
(15), (16) in [22]). It  exhibits the property of both confinement
and chiral symmetry breaking.  The states are in general
characterized by the quantum numbers $j,l,\kappa$. In particular,
the $D$ and the orbitally  excited state $D_s^{*}$ correspond to the
solution of the ground state in the $j,l,\kappa=1/2,0,-1$ and
$j,l,\kappa=1/2,1,+1$ channel respectively.

In this study a value of $\sigma=0.18$~GeV$^2$ and $\alpha_s=0.35$
is adopted. For convenience, a zero mass is used for the $u,d$
quark, while for the $s$-quark we have taken $m_s=200$~MeV. Having
constructed these wave functions with these parameters we
determine the matrix elements $v_{12}^{(K)}$. From this we may
then solve the resulting Eq. (\ref{33}) iteratively. To determine
the actual position of the pole we in general have to analytic
continue the integral into the second sheet in the case that $E$
is above the $KD$-threshold. Although this can be done, we will
assume in this study that the imaginary part of the pole position
does not affect the solution  substantially , which is certainly
true when the pole dives under the $DK$-threshold, and was checked
in other situations. Confining ourself to real values of E and
taking the principal value of the integral in Eq. (33) when E is
above threshold the energy shift is determined as a solution of
the resulting equation. In case of a solution above threshold, the
width of the resonance can be obtained by calculating the
discontinuity of the integral at this energy.

In the calculations we have used for the threshold mass of the KD
system $m_K+m_D=2.366$~GeV. A typical value of $2.49$~GeV is
adopted for the unperturbed $D_s^{*}$ meson mass. As is seen in
Table 3 this is in accordance with the predictions of FCM and of
many quark constituent quark models.

\begin{figure}
\begin{center}
\includegraphics[width=10cm]{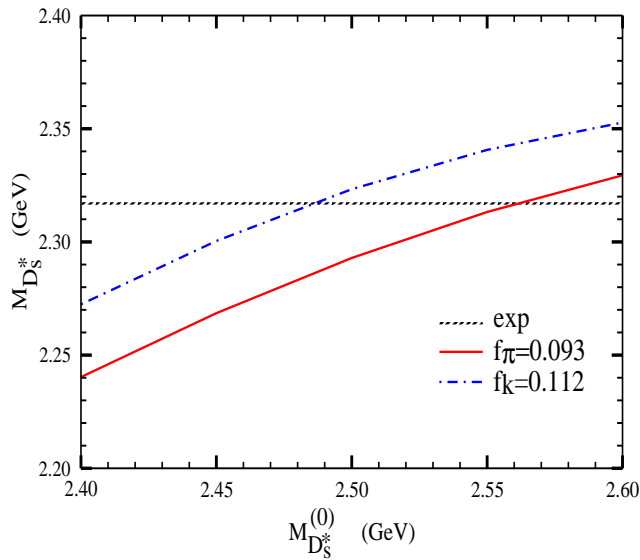}
\caption{ The shifted $D_s^{*}$ meson  mass as a function
of the unperturbed $M^{(0)}_{D_s^{*}}$ for two values of the 
decay constant $f$. The experimentally
observed mass is given by the dashed horizontal line}
\hfill
\label{fig2}
\end{center}
\end{figure}

In Fig. 2 is shown the prediction of the shifted mass of the
$D_s^{*}$ meson due to the channel coupling  to the $KD$-system as
a function of the unperturbed mass $D^{*(0)}_{s}$. The flavoured
symmetric and broken value of $f_\pi=93$~MeV and $f_K=121$~MeV
have been used [1]. The shifted mass is found using
\be
M_{D^{*}_s} = M_{D^{*(0)}_{s}}+\Delta E,
\ee
where $\Delta E$
satisfies Eq.~(\ref{33}). Due to the coupling to the $KD$ system
the $D_s^{*}$ meson can either become unstable or stable,
depending on the sign and magnitude of the mass shift. We find in
our  model, that it can be substantial and of the order of hundred
MeV. As is seen from the figure we find the experimentally
observed value of $2.317$~GeV for an unperturbed mass of
approximately  $2.49$~GeV.

In the string model studied we find that the position of the pole
moves well down below the $KD$ threshold, yielding a stable state.
  The channel coupling is found to yield
attraction, so that the position of the mass pole shifts downwards.

Similarly as for the $D^*_s(2317)$ meson we may estimate the mass
shift of the $D^*_s(2632)$ due to channel coupling. The calculations
proceeds in the same way. The bare $D^*_s(2632)$ is assumed to
correspond to the first radial excited state with
$j,l,\kappa=1/2,0,-1$ and to have a mass of $2.76$ GeV as found in
the FCM \cite{13}. Clearly a mass shift of about $-140$~MeV is
needed to get agreement with the observed mass. As discussed,
coupling can occur in this case to $KD$ and $\eta D_s$ channels. For
the $D$ meson we assume the experimental observed mass of
$1.869$~GeV. There are two candidates for the $D_s$ state, being the
$0^-$ and $1^-$ states with masses of $1.968$ and $2.120~GeV$
respectively. Using the various wave functions obtained from the QCD
string model, the interaction matrix elements are calculated from
Eq.~(\ref{29}). The two $D_s$ states are degenerate in the
considered model and as a result have the same wave function, but
differ in the kinematics of the momenta in view of the adopted mass
difference, which occurs due to the hyperfine interaction neglected
in our heavy-quark approximation. For simplicity we have considered
only one $D_s$ state with mass $2.0~GeV$. With the obtained
potential matrix elements we solve numerically the eigenvalue
equation for $\Delta E$. It has essentially the same form as
Eq.~(\ref{33}), but has now two terms due to the contributions from
the two coupled channels.

We find that the bare mass is shifted downwards by $51~MeV$.
Furthermore, the contributions from the various inelastic channels
are given by $$\Delta E(KD)=-35~{\rm MeV},~\Delta E(\eta D_s)=-16~
{\rm MeV}.$$  The magnitude of the total mass shift is clearly
smaller here than found in the first considered case of the
$D_s^*(2317)$ meson as can  be explained by the presence of
$P$-wave, rather than $S$-wave for the case of $D^*_s(2317)$. It
is clearly not sufficient to explain the experimental observed
mass. We can also calculate the width, being the discontinuity of
the right hand side of Eq.~(33). We find $\Gamma=174$ MeV,
decaying predominantly into the $K D$-channel. The
corresponding partial decay widths to the various channels are
found to be $$\Gamma(KD)=139~{\rm MeV},~\Gamma(\eta D_s)=35~{\rm
MeV}.$$

Clearly the predicted width is considerably larger than found in
the SELEX experiment. We have also solved Eq.~(33) for the case of
the  flavour symmetric value $f_K=f_\eta=.093~GeV$, which yields
again a large value for the width.  The above results suggest in
this case that the resonance becomes very broad and does not
support the SELEX observation,  also the ratio of $KD$ to $\eta
D_s$ channels quantitatively disagrees with the experiment.

In general, the size of the mass shift clearly depends on
structure of the quark wave functions and hence it should be
expected to be model dependent. Our study demonstrates, that the
size of the shift due to channel coupling is in general large, but
that it also can lead to very large widths in case that the
resonance is above threshold of the coupled channels. As a result
it can accommodate for the discrepancy between the predictions of
dynamical quark models and the observed $D^*_s(1237)$ resonance,
but there may exist situations where the resonance can become very
broad due to inelastic channel coupling, as it is in the case of
radial excited $D^*_s$ .

\section{Discussion}

 Let us compare our results to the existing  in literature. The
explanation of $D_s^*(2317)$ as the $c\bar s$ $p$-wave level
shifted down by the coupling to the decay channel $D+K$ was
considered in a series of papers [19], where authors have used
 a simple phenomenological model similar to our Eq. (26), (27) to
 describe the $p$-wave mesons.

 Another type of phenomenological model for the channel coupling,
 namely the model of Eichten at al. [26] was used in Ref.~\cite{27} to
 calculate the shift of the $ D_s (0^+)$ level and it was shown that
 the desired mass shift is obtained for a reasonable choice of
 parameters.

 Our results obtained with the parameter free chiral Lagrangian
 containing full $x$-dependence,  qualitatively agree with those
 in Refs.~[19] and [27] and  exactly reproduce the experimentally found
 mass $D^*_s(2318).$

  We now turn to the state $D_s^*(2632)$, found in Ref.~\cite{14},
but not yet confirmed by other groups \cite{28}. The theoretical
prediction for the $2^3S_1$ state vary from 2774 MeV in
Ref.~\cite{13} to 2737 MeV made using the relativistic Salpeter
equation in Ref.~\cite{29} and 2716 in Ref.~\cite{8}, and in
principle are subject to the correction due to the global string
breaking effect occurring for states of large size \cite{30} (m.s.r.
radius of $2^3S_1$   state in Ref.~\cite{29} is around 1 fm). The
expected correction is around -- 20 MeV, which brings the
theoretical mass of the $D_s(2^3S_1)$ state  to 2700-2720 MeV.

Our calculation  for the initial mass $D^*_s(1^-),~~ m=2710$ MeV,
using Eq. (33), yields the shift $\Delta E=-76$ MeV, with the
total width $\Gamma=131$ MeV, $ \Gamma_{KD} =111$ MeV,
$\Gamma_{D_s\eta}=20$ MeV. Hence also in this case the total width
and width ratio contradicts experimental data, implying that
$D^*_s(2632)$ cannot be  explained as the  shifted  $1^-$ level.

\section{Acknowledgements}
Yu.A.S. acknowledges useful discussions with A.M.~Badalian and J.A.T.
is grateful to J.~Goity for a useful discussion. The work of J.A.T.
is supported by the U.S. Department of Energy contract number
DE-AC05-84ER40150 under which the Southeastern Universities Research
Association (SURA) operates the Thomas Jefferson National
Accelerator Facility. The work Yu.A.S. is supported by the Federal
Program of the Russian Ministry of Industry, Science, and
Technology No 40.052.1.1.1112.\\

\end{document}